\documentclass[12pt]{article}

\usepackage[pdftex]{graphicx}

\begin{document}

\title{The principle of a finite density of information}
\author{Pablo Arrighi\thanks{
\'Ecole normale sup\'erieure de Lyon, LIP, 46 all\'ee d'Italie, 69008 Lyon, France,
and
Universit\'e de Grenoble, LIG, 220 rue de la chimie, 38400 SMH, France,
{\tt parrighi@imag.fr}.}
~and~Gilles Dowek\thanks{INRIA, 23 avenue d'Italie, CS 81321, 75214 Paris Cedex 13, France,
{\tt gilles.dowek@inria.fr}.}}
\date{}
\maketitle
\thispagestyle{empty}

\begin{abstract}
The possibility to describe the laws of the Universe in a
computational way seems to be correlated to a principle that the
density of information is bounded. This principle, that is dual to
that of a finite velocity of information, has already been
investigated in Physics, and is correlated to the old idea that there
is no way to know a magnitude with an infinite precision. It takes
different forms in classical Physics and in quantum Physics.
\end{abstract}

\section{Why cellular automata?}

Stephen Wolfram has advocated in \cite{NKS} the idea that cellular
automata might be a relevant formalism to study the laws of the
Universe. This thesis can be seen as a consequence of three more
fundamental principles that are implicit in the definition of the
notion of a cellular automaton.

To define a cellular automaton, we must discretize space by
partitioning it into an infinite set of identical {\em cells}, for
instance cubes.  We must also discretize time and
observe the Universe at regular time steps.  Then come three
assumptions
\begin{enumerate}
\item that the state space of each cell is finite,
\item that the state space of a cell at a given time step depends only on 
the state of a finite number of neighbours at the previous time step,
\item that the state space is the same for all cells and the evolution acts the same everywhere and everywhen.
\end{enumerate}

Assumption (3.) is a reformulation of a well-known principle of
Physics: the homogeneity of space and time. So is assumption (2.),
which is a reformulation of the bounded velocity of information.
Assumption (1.), in contrast, seems to express the new idea, that the
density of information also is bounded. Such a principle can be
formulated as the fact that the amount of information that can be
stored in a bounded region of space is bounded. It can also be
formulated as the fact that the cardinality of the state space of a
bounded region of space is bounded, since the amount of information
storage is the logarithm of this cardinality.

There are other assumptions in the definition of a
cellular automaton, such as the fact that space-time is absolute and
passive: the neighbourhood relation on cells does not evolve as their
states do. Weakening these assumptions is possible, leading for instance to
{\em causal networks} \cite{NKS} or to
{\em causal graph dynamics} \cite{ArrighiCGD}.

There is a historical discrepancy between the idea of a finite
velocity of information and that of a finite density of
information. The first has had a clear status as a principle of
Physics, since special relativity, and we even know the bound on the
velocity. The second seems to be less established. Yet, it is not
completely new, as all three principles (1.), (2.), and (3.), have
been stated by Robin Gandy in \cite{Gandy}, and the principle of a
bounded density of information has also been stated by Jacob
Bekenstein---although the Bekenstein bound is not on the amount of
information but on the quotient of the amount of information and the
energy contained in a sphere \cite{Bekenstein} .

\section{Two dual principles}

The principles of a bounded velocity and of a bounded density of
information play a dual role.

For instance, the amount $I$ of information that can be output by a
wire of section $S$ in a given time $t$ is bounded by the amount of
information contained in the cylinder of section $S$ and length $l$,
where $l$ is the distance that information can travel in time $t$.

\begin{figure}[h]
\centering
\includegraphics[scale=1.0, clip=true, trim=0cm 0cm 0cm 0cm]{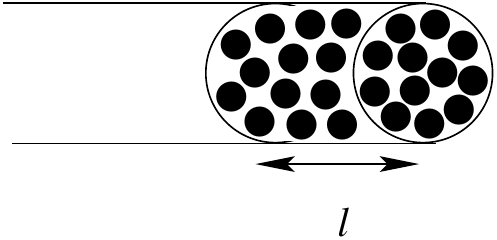}
\end{figure}

If information has a bounded velocity, then $l$ is bounded and if
moreover information has a finite density, the amount of information
$I$ is bounded. Thus we need both principles to prove that the
information flow of a wire is bounded. If one of these principles were
invalid, it would be easy to build a wire with unbounded flow.

A more evolved example is the impossibility to build a infinitely
parallel computing machine, i.e. a machine formed with an infinite
number of processors communicating with a common server. If
information has a bounded density, then the size of each machine has a
lower bound and some machines must be arbitrarily far from the server
and if moreover information has a finite velocity, some machines are
too far to communicate with the server in due time. Again, if one of
these principles were invalidated it would be easy to build an
infinitely parallel machine, either with processors of the same size
communicating infinitely fast with the server,
\begin{figure}[h]
\centering
\includegraphics[scale=0.5, clip=true, trim=0cm 0cm 0cm 0cm]{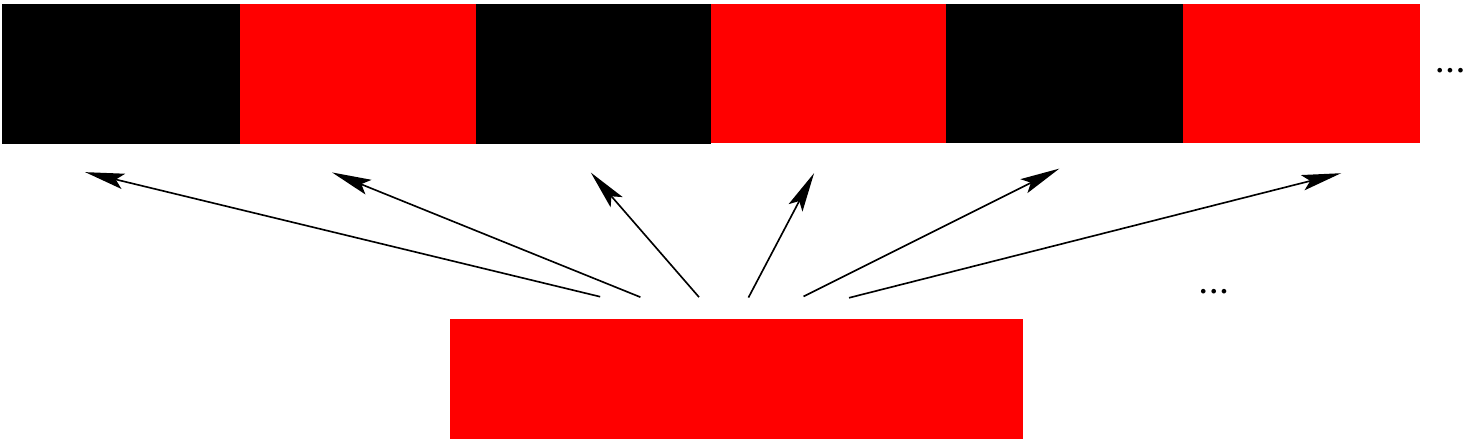}
\end{figure}
or with arbitrarily small machines, of size 1, 1/2, 1/4, 1/8, 1/16,
\ldots that would lie at a finite distance of the server.

\begin{figure}[h]
\centering
\includegraphics[scale=0.5, clip=true, trim=0cm 0cm 0cm 0cm]{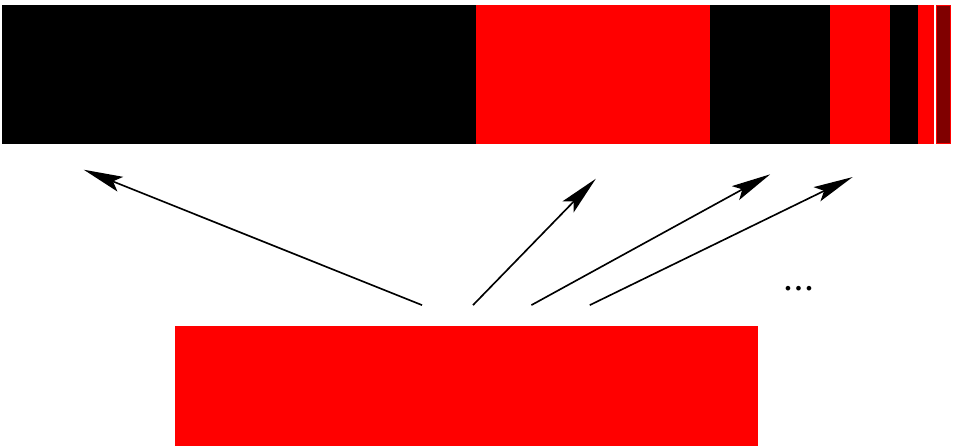}
\end{figure}

It seems that these two principles could even be summarized in one, as
the fact that in a given time, a piece of information can only travel
in a space which is itself populated by a finite amount of
information.

\begin{figure}[h]
\centering
\includegraphics[scale=0.5, clip=true, trim=0cm 0cm 0cm 0cm]{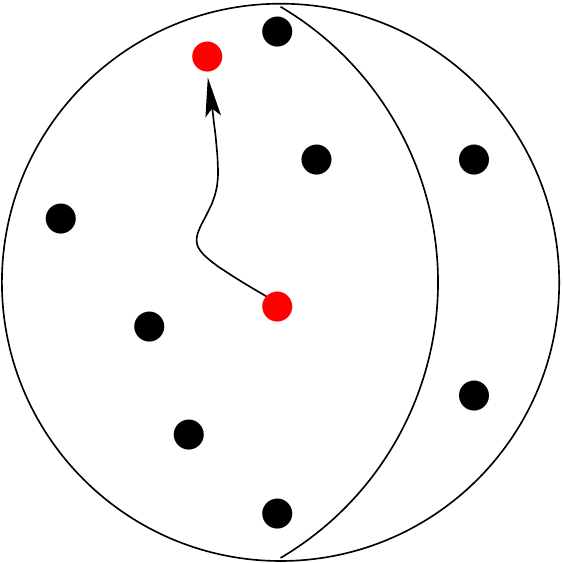}
\end{figure}

Finally, the two principles are comparable, and both differ from the
third: homogeneity of space and time. Indeed, the latter is
pre-supposed by the very notion of a `Physics law'. For instance,
suppose that some law would be valid in Paris but not in Boston, then
this would not be much of a law. The word `law' itself carries a
notion of universality.

But what about the other two? It turns out that a it is possible to do
very respectable Physics without these principles. Yet, it is still
the case a notion of applicability of a Physics laws underpins both
the finite velocity of information and the bounded density of
information principles. If a law broke the finite velocity of
information principle, this would entail that, in order to determine
the new state of a physical system after one second has past, one may
potentially need to look at an unbounded number of neighbours. This is
no doubt unpractical, unless far-away contributors can be neglected.

In the same way, if a law broke the finite density of information
principle, in order to determine the new state of a physical system,
one would need to use a description of the state of the neighbours
that might potentially contain an unbounded amount of
information. Again this law would hardly be applicable, unless this
information can be rounded up in some way.

\section{Newtonian Physics}

It is well-know that Newtonian Physics contradicts the principle of a
finite velocity of information. Whenever a bee flies half a
centimeter, it instantaneouly modifies the gravitational field on a
distant galaxy. It is only with General Relativity that we understood
how to control the consequences of the flight of a bee.

Newtonian Physics not only contradicts the finite velocity principle,
it also contradicts the principle of a finite density of information
as the distance between the jaws of a caliper for instance is measured
by a real number. So if we imagine an infinite amount of information
expressed, for instance, as an infinite sequence of 0 and 1, we can
see this infinite sequence as the digits of a real number and place
the jaws of a caliper at this exact distance, recording this way the
infinite amount of information in a finite volume.

Of course, we have all been taught that this idea is nonsense and that
the distance between the jaws of a caliper is defined with a finite
precision: three or four digits. It is surprising that this thesis
that a physical magnitude is always defined with a finite precision
has never been given the status of a Physics principle as have the
homogeneity of space and time. Yet, this thesis may be a very early
occurrence of this idea of a finite density of information.

\section{The finite-density versus the superposition principle}

The principle of a finite density of information seems to be
challenged by quantum theory and its superposition principle. Indeed,
regardless of whether a region of space is bounded or not, whenever
its state space contains two states ${\bf u}$ and ${\bf v}$, then it
must also contain all the linear combinations $\lambda {\bf u} + \mu
{\bf v}$ up to a renormalization factor.  Hence it is infinite, and
the finite density principle cannot be formulated as the fact that the
set of states of a bounded region of space is finite.

These complex amplitudes $\lambda$ and $\mu$ can be compared with
probabilities. But this comparison has its limits. Indeed, since
probabilities are positive real numbers, they can only add up
constructively to make and event more likely. In contrast amplitudes
are complex numbers, then can also be substracted to one another,
i.e. add up destructively, making an event either more likely of less
likely; or sometimes even impossible.

This is why when two different scenarios happen in a superposition, we
cannot quite think of them as two non-interacting branches of what may
happen in the Universe: these amplitudes cannot be ignored. 
This makes the superposition principle is a difficulty when
extending the finite-density principle to quantum theory.

Yet, the infinity of the state space does not mean that the amount of
possible outcomes, when measuring the state of this region, is itself
infinite, because in quantum theory, we cannot measure the state of
the system, but only one of its observables.  Thus, an alternative
formulation of the bounded density of information principle is that
each projective measurement of a bounded region, at any given point in
time, may only yield a finite number of possible outcomes. This
requirement amounts to the fact that the state space of a bounded
region of space is a finite-dimensional vector space.  This
constitutes a good quantum alternative to the formulation of the
finite density of information principle, one which does not demand
anything to be actually measured in any way.

\section{The finite-density principle versus correlations}

Moreover, yet another problem arises in the quantum setting, mamely
that of `correlations'; which in this context are also referred to as
`entanglement'. To understand what entanglement is, one must apply the
superposition principle again, but this time to pairs of systems. For
instance if systems $A$ and $B$ may be in state ${\bf u u}$, meaning
that both are in state ${\bf u}$, or in state ${\bf v v}$, meaning
that both are in state ${\bf v}$, then the state $\lambda ({\bf
  u u})+\mu ({\bf v v})$ is also allowed. This entangled state
corresponds to a superposition of two global, and everywhere different
scenarios. It is possible to get an intuition of the meaning of this
entangled state by appealing to our understanding about usual
probabilities. In probability theory, it is a commonplace that knowing
the marginal probabilities over the individual systems does not entail
knowing the joint probability distributions. For instance Alice and
Bob may have a half-half chance of having a blue or red ball in each
of their boxes, but it could be that each of their boxes have been
prepared in such a way that both balls are of the same colour. The
state $\lambda ({\bf u u})+\mu ({\bf v v})$ tells the same story not
for probabilities, but for amplitudes.

This is an issue, because the bounded velocity of information
principle seems very weak in that respect: after all, all it says is
that the state of each individual system is determined by that of the
neighbours, but it does not say how the entanglement between the
individual states is determined. If we retake the probability analogy;
the bounded velocity of information principle tells you that the
probabilities of Alice finding a red or a blue ball in her box are
determined locally; but what about the probability that this ball is
of the same color as Bob's?

Fortunately on this issue of whether the correlations can be
determined locally, quantum theory is a bit friendlier than
probability theory. Indeed, it has been shown that whereas the bounded
velocity of information principle is way too weak in the context of
probabilistic cellular automata \cite{ArrighiPCA}, surprisingly it
turns out to be sufficient in the context of quantum cellular automata
\cite{ArrighiJCSS} for the evolution to be described as a composition
of local functions mapping neighbourhoods to neighbourhoods, that is
functions from a finite-dimensional vector space to itself. Another
way to phrase this is that in the quantum setting, any evolution that
respects the bounded velocity of information principle necessarily
takes the form of a circuit of gates being applied on systems and
their neighbours. Thus correlations can be `tamed' and the global
evolution broken into functions from finite-dimensional vector spaces
to themselves.

\section{A computable quantum Universe}

Yet, moving from a finite set to a vector space, even a
finite-dimensional one, is a big leap and many advantages of
describing the Universe as a cellular automaton might be lost by doing
so. Indeed, the main advantage of having a finite state space $S$ for
cells was that the local evolution function of the cellular automaton,
that is a function from $S^n$ to $S$, where $n$ is the number of
neighbours of a cell, was a function from a finite set to a finite
set. Such a function can be described by a table giving the image of
each element of the domain and it is always computable. This point
was the key to the computability of cellular automata, and thus to the
idea that the Universe is computable if it can be described as a
cellular automaton. This idea of a computable Universe---the physical
Church-Turing thesis---is one of the main goals of both Robin Gandy and
Stephen Wolfram.

Fortunately, another feature of the quantum theory comes to repair the
damages created by the superposition principle. According to the
quantum theory, theses local functions of a quantum cellular automata
cannot be any function from neighbourhoods to neighbourhoods, but must
be linear. There are much less linear functions than functions from a
finite-dimensional vector space to another. In particular any such
linear function can be described by a matrix, and matrix
multiplication boils down to additions and multiplications that are
computable operations. Thus if scalars are chosen in an adequate
manner as discussed below, all such functions are computable. Thus,
infinity is tamed by finite-dimension and linearity and this
formulation of the principle of a finite density of information is
sufficient to prove the Universe computable.

\section{Scalars}

When saying that addition and multiplication of scalars are
computable, we need to be a little careful, as scalars are complex
numbers and computability is mostly defined for natural numbers.

Computability can be extended from natural numbers to real and complex
numbers \cite{Weihrauch}, but then the picture is a somewhat
different.  Not only functions from complex numbers to complex numbers
can be computable or not, but complex numbers themselves can be
computable or not. Thus, although multiplication is computable,
multiplication by a fixed scalar may be non-computable if the fixed
scalar is non-computable.

Michael Nielsen has noticed that having a process building a superposed state
$\lambda {\bf u} + \mu {\bf v}$, where $\lambda$ and $\mu$ are non
computable numbers, can {\em per se} lead to non computability, as this
state can be tomographed by repeated measurements and $\lambda$ be
extracted with increased precision from the state.

Thus, a physical system has two ways to encode an infinite amount of
information: either by having an infinite dimensional state space, or
by encoding an infinite amount of information in a scalar, exactly in
the same way as an infinite amount of information can be encoded in a
distance in Newtonian Physics.

To express the finite-density principle, it seems that it not
sufficient to restrict to finite-dimensional vector spaces for the
state spaces of cells, but we must also restrict to a smaller set of
scalars, yielding an intersting problem for physicists about the
fields that can be used for the scalars in quantum theory.

With this precise formulation of the finite density of information, we
can formally prove that the physical Church-Turing thesis is a
consequence of the three hypotheses we have mentioned: homogeneity of
space and time, bounded velocity of information and bounded density of
information \cite{ArrighiGANDY}.

Some scientists, such as David Deutsch \cite{Deutsch} propose to take
the physical Church-Turing thesis as a principle of Physics. Some
others, such as Michael Nielsen \cite{NielsenComputability}, propose
to deduce this thesis from more fundamental principles of
Physics. Like Robin Gandy, we have chosen to start from these three
fundamental principles.  An advantage of this approach is that these
three principles are falsifiable provided explicit bounds are given.
Thus, from a logical point of view, they are purely universal
formulas, while the physical Church-Turing thesis itself has a much
more complex logical form $\forall \exists \forall$, namely for all
physical system $m$, there exists a computable function $f$ such that
forall initial state $x$, the evolution of $m$ leads from $x$ to
$f(x)$.

\section{Conclusion: cellular automata and quantum cellular automata}

It is one thing to say that the Universe is computable and thus can be
simulated by any model of computation, e.g Cellular Automata. It is
another to seek to closely model physical systems in a
space-preserving manner. Pursuing this second, more demanding aim
leads us to investigate how to model the state space of a bounded
region of space. Supposing that this state space is finite is adequate
in the classical setting, but must be refined in the quantum setting
to a finite-dimensional vector space over a reasonable field. Indeed
it seems that simulating a quantum systems by classical cellular
automata not only leads to an exponential slowdown, but also fails to
respect the geometry of the system. This justifies Feynman's proposal
to model quantum systems with Quantum Cellular Automata.  This quantum
extension of the Cellular Automata model shows its liveliness and
robustness.

\end{document}